\documentclass[%
 aip,
 sd,%
 amsmath,amssymb,
 reprint,%
]{revtex4-2}

\usepackage{multirow}
\usepackage{graphicx}
\usepackage{bm}
\usepackage{hyperref}

\usepackage{graphicx, textcomp, amssymb, mathrsfs}

\makeatletter
\newcommand\xleftrightarrow[2][]{%
\ext@arrow 9999{\longleftrightarrowfill@}{#1}{#2}}
\newcommand\longleftrightarrowfill@{%
\arrowfill@\leftarrow\relbar\rightarrow}
\makeatother

\usepackage{epstopdf}

\begin{document}

\title{On the behavior of a distributed network of  capacitive constant phase elements}

\author{Anis Allagui$^*$}
\email{aallagui@sharjah.ac.ae}
\affiliation{Dept. of Sustainable and Renewable Energy Engineering, University of Sharjah, Sharjah, P.O. Box 27272, United Arab Emirates}
\altaffiliation[Also at ]{Center for Advanced Materials Research, Research Institute of Sciences and Engineering, University of Sharjah, Sharjah, P.O. Box 27272,  United Arab Emirates}
\affiliation{Dept. of Electrical and Computer Engineering, Florida International University, Miami, FL33174, United States}

\author{Ahmed Elwakil} 
\affiliation{
Dept. of Electrical Engineering, 
University of Sharjah, PO Box 27272, Sharjah, United Arab Emirates 
}

\begin{abstract}

As a generalization of integer-order calculus,   fractional calculus has seen tremendous applications in the past few years especially in the description of anomalous viscoelastic properties, transport processes in complex media as well as in dielectric and impedance spectroscopy of materials and electrode/electrolyte interfaces. The fractional-order capacitor or  constant phase element (CPE) is a fractional-order  model  with impedance $z_c(s) = 1/(C_{\alpha} s^{\alpha})$ ($s=j \omega$, $C_{\alpha}>0$,  $0<\alpha<1$) and is widely used in modeling impedance spectroscopy data in dispersive materials. In this study, we investigate the behavior of a network  of distributed-order CPEs, each of which described by a Caputo-type time fractional differential equation relating the current on the CPE to its voltage, but with a non-negative, time-invariant weight function $\phi(\alpha)$. The  behavior of the distributed-order network in terms of impedance   and time-domain response to a constant current excitation is derived for two simple cases of $\phi(\alpha)$: (i) $\phi(\alpha)=1$ for $0 < \alpha < 1$ and zero otherwise, and $(ii)$ $\phi(\alpha) = \sum_i C_{\alpha_i}\, \delta(\alpha-\alpha_i)$  corresponding to the general case of parallel-connected elemental CPEs of different orders $\alpha$, and pseudocapacitances $C_{\alpha}$. Our results show that the overall network is not equivalent to a single CPE, contrary to what  would of been expected with ideal capacitors. \\

Keywords: Fractional-order capacitor; Constant phase element;  Impedance spectroscopy; Fractional calculus. 
 
\end{abstract}

\maketitle

\section{Introduction}

A constant phase element (CPE) represents a fractional-order capacitive-resistive impedance model that fills the gap between the behavior of the ideal capacitor and that of the ideal resistor\;\cite{amani2015analysis}. Its dispersive lossy nature is usually attributed   to distributed surface reactivity and surface inhomogeneity and as a result to current and potential distributions, and also to the roughness, fractal structure and porosity of the electrode \cite{jorcin2006cpe}.
 From a modeling perspective, the CPE is regarded as a standard tool for the analysis and description of non-ideal dielectric and impedance spectroscopy data of real capacitive systems\;\cite{gateman2022use, orazem2013dielectric, vendrell2022revealing, abouzari2009physical}. The CPE behavior  in the time-domain in response to different types and forms of excitations has been investigated for example in \;\cite{allagui2024exact, 10.1149/1945-7111/ac621e, allagui2021inverse, ragone}. In nearly all applications in which the CPE is invoked, the order $\alpha$ (also known as the dispersion coefficient) of the CPE   is usually taken as a   constant and single-valued parameter, independent of any other implicit variables involved in the dynamics of the system under study\;\cite{lasia2022origin}. 

However, because of the ever-increasing  complexity   of the advanced electrodes used today, the fact that they may be of heterogeneous composition, and the possible interplay between multiple co-existing orders (associated with sub-domains in the system),  it is important and natural  to consider a more general approach of a distributed network of CPE with multi-valued pseudocapacitances and orders \;\cite{caputo2001distributed, ding2021applications, patnaik2020applications}. Distributed order fractional dynamics   implies variations   of the order with some variables, but not with time \;\cite{bagley2000existence, bagley2000existence2} (e.g. variations with temperature). Whereas studying the dynamics of   complex systems while taking into account variations of the order with respect to time is referred to as variable-order fractional dynamics\;\cite{patnaik2020applications}, and is beyond the scope of this work. 
This work is on the generalization of the CPE using the class of distributed order operators  based on the Caputo fractional time derivative relating the CPE's current and voltage. We focus on the capacitive CPE, knowing that it should straightforward to mirror the whole study to the case of inductive CPE. Specifically, we derive the frequency-domain impedance of the CPE network, and its time-domain  expressions for the voltage in response to constant current excitation in closed forms using the Mittag-Leffler and Fox's $H$-function properties\;\cite{haubold2011mittag, mathai2009h, mathai2008mittag, mathai1978h}. This is done for two cases of (i) a uniformly distributed function for the order $\alpha$, and (ii) a set of discrete values of $\alpha$, which can be viewed as reliable models for real electrode systems.

\section{The constant phase element}

We first recall the CPE's current-voltage constitutive equation, which is given by \cite{allagui2025semi}: 
\begin{equation}
i_c(t) = C_{\alpha}\,
{}_0\text{D}_t^{\alpha} v_c(t)
\label{eq:iCPE}
\end{equation}  
where the pseudocapacitance $C_{\alpha}$ is in units of F\,s$^{\alpha-1}$, and    
 ${}_0\text{D}_t^{\alpha} $ is the Caputo differential  operator of constant order $\alpha$ ($0<\alpha < 1$) defined as:
\begin{equation}
{}_0\text{D}_t^{\alpha} f(t) := \frac{1}{\Gamma(m-\alpha)} \int_0^t (t-\tau)^{m-\alpha-1} f^{(m)}(\tau) d\tau
\end{equation}
Here $m\in \mathbb{N}$, $m-1< \alpha < m$ ($m=1$ in our case), $\Gamma(z) =\int_0^{\infty} u^{z-1} e^{-u} du,\,(\text{Re}(z)>0)$ is the gamma function, and $f^{(m)}(t)=d^m f(t)/dt^m$ is the $m^{th}$ derivative of $f(t)$ with respect to\;$t$. 
Knowing that the Laplace transform  of the Caputo  fractional derivative of order $\alpha$ is\;\cite{mathai2009h}:
\begin{align}
\mathcal{L}\left[{}_0\text{D}_t^{\alpha} f(t); s \right] \nonumber
 &= \int_0^{\infty} e^{-st} {}_0\text{D}_t^{\alpha} f(t)\, dt \\
 &= s^{\alpha} \tilde{f}(s) - \sum\limits_{k=0}^{m-1} s^{\alpha-k-1} f^{(k)}(0^+)
\end{align}
and assuming zero initial conditions ($f^{(k)}(0^+)=0$), it is straightforward to obtain the frequency-domain impedance of the CPE as the ratio of the Laplace transform of the time-domain voltage   by that of the time-domain current   as \cite{10.1149/1945-7111/ac621e, allagui2021possibility, allagui2021inverse,cpe}:
\begin{equation}
z_c(s)
 = \frac{\tilde{v}_c(s)}{\tilde{i}_c(s)} =  \frac{1}{C_{\alpha} s^{\alpha}}
\end{equation} 
The phase is $\phi(z_c)=\tan^{-1}(-\alpha \pi/2)$, constant and independent from frequency.  The real part of the CPE impedance is  $\text{Re}(z_c) = \cos(\alpha \pi/2)/(C_{\alpha} \omega^{\alpha})$, and its imaginary part is   $\text{Im}(z_c) = -\sin(\alpha \pi/2)/(C_{\alpha} \omega^{\alpha})$. 

In the case that a current $i_c(t)$ is applied to a CPE,  the developed voltage $v_c(t)$ can be found by applying  the inverse Laplace transform using the convolution theorem as:
\begin{equation}
v(t)
=\mathcal{L}^{-1}\left[ \tilde{z}_c(s)  \,{\tilde{i}_c(s)}  ; t \right]
= \frac{1}{C_{\alpha} \Gamma(\alpha)} \int_0^t (t-\tau)^{\alpha-1} i_c(\tau) d\tau
\end{equation}
The same can be carried out if a voltage-excitation is applied and the current is to be calculated \cite{allagui2024exact, allagui2023tikhonov}. 

In the following section, we proceed to study the behavior of a network of distributed-order CPEs according to a certain distribution function.


\section{Distributed-order constant phase element network}

\subsection{Theory}

Now we consider the following distributed-order time differential equation\;\cite{ding2021applications, lorenzo2002variable, eab2011fractional, bagley2000existence, bagley2000existence2} as a describing equation for a CPE network (compare with Eq.\;\ref{eq:iCPE}):
\begin{equation}
i(t) = 
\int_{\beta_1}^{\beta_2} \phi(\alpha)\, C_{\alpha}\,
{}_0\text{D}_t^{\alpha} v_c(t) d\alpha
\label{eq:iCPE2}
\end{equation}
Here $\phi(\alpha)$ is a non-negative, time-invariant weight function defined on the interval $[\beta_1, \beta_2]$ for $\alpha$, and acts as a (discrete or continuous) distribution of orders. 
The  integral in Eq.\;\ref{eq:iCPE2} is known as a cumulative order distribution over the range of orders $\beta_1 \leqslant \alpha \leqslant \beta_2$\;\cite{lorenzo2002variable}, and is a direct generalization of the constant-order Caputo fractional derivative.  
For our purpose, we will restrict ourselves to the lower and upper values  of the range of integration to zero and one (i.e. $0 < \beta_1 \leqslant \alpha \leqslant \beta_2 < 1$), which defines the spectrum going from an impedance of a pure resistor  ($\alpha=0$) to that of  a fractional capacitor ($0 < \alpha < 1$),  to that of an ideal capacitor ($\alpha=1$). 
It is worth noting that Eq. \;\ref{eq:iCPE2}  is one possible generalization of Eq.\;\ref{eq:iCPE}, amongst others, which can be used here to describe the current-voltage dynamics of more complex systems than a single-order CPE, such as heterogeneous electrode/electrolyte interfaces with multi-fractal properties\;\cite{eab2011fractional}.  
A similar equation was studied for instance by Eab and Lim\;\cite{eab2011fractional} for the case of simple free fractional Langevin equation without a friction  term, i.e. with the left-hand side of Eq.\;\ref{eq:iCPE2} being a stationary Gaussian random noise. 
   Lorenzo and Hartley also studied a similar problem of distributed order operators in the context of  viscoelastic materials\;\cite{lorenzo2002variable}. 

By applying the Laplace transform to Eq.\;\ref{eq:iCPE2},   with the   assumption of zero initial conditions, we obtain:
\begin{equation}
\tilde{i}(s) =   \tilde{v}_c(s) \int_{\beta_1}^{\beta_2} \phi (\alpha)\, C_{\alpha} s^{\alpha} d \alpha
\label{eq:iLaplace}
\end{equation}
The corresponding impedance function is therefore:
\begin{equation}
\tilde{z}(s)= \left( {\int_{\beta_1}^{\beta_2} \phi (\alpha)\, C_{\alpha} s^{\alpha}  d \alpha} \right)^{-1}
\label{eq:zofs}
\end{equation}
and the time-domain voltage $v(t)$ can be found     by the convolution integral:
\begin{equation}
v_c(t)=  \int_0^t z(t-\tau) i(\tau) d\tau
\end{equation}
where
$z(t) = \mathcal{L}^{-1}\left[\tilde{z}(s) ; t \right]$.

The specific choice of $\phi(\alpha)$ in Eq.\;\ref{eq:iCPE2} depends on the underlying physics of the system under study.  
In what follows, we focus on two examples that can describe close enough  what one may encounter when analyzing heterogeneous, complex electrodes in electrolytic media. 

\subsection{Example 1: $\phi(\alpha)$ is a uniform distribution function}

\begin{figure}[!t]
\begin{center}
\includegraphics[angle=-90,width=0.465\textwidth]{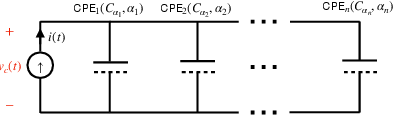}
\caption{Current-excited parallel network of  CPEs with distributed parameters $C_{\alpha_i}$ and $\alpha_i$}
\label{fig1}
\end{center}
\end{figure}

We first consider  the case of a uniformly distributed order, i.e. a weight function $\phi(\alpha)$  given by:
\begin{equation}
\phi_u(\alpha) = 
\begin{cases}
     1,\; 0 \leqslant \alpha \leqslant 1  \\
     0,\; \text{otherwise}           
\end{cases}
\label{WH}
\end{equation}
This means that the network (see Fig.\;\ref{fig1}) is composed of many elemental CPEs of different orders (taken from  zero to one) while the value of $C_{\alpha}$ is the same for all  CPEs; i.e.  $C_{\alpha_1}=C_{\alpha_2}=\ldots=C_{\alpha_n} = C_{\alpha}$). The contributions of each and every elemental CPE are  equally identical. Note that $\int_0^1 \phi(\alpha) d\alpha$ does not need to be necessarily equal to one in this example, or in the general definition given above in Eq.\;\ref{eq:iCPE2}\;\cite{mainardi2008time}. 
 
  With the above assumptions, we obtain  the overall network equivalent impedance as:
\begin{align}
\tilde{z}_u(s) &= \left( C_{\alpha} {\int_{0}^{1}  s^{\alpha} d \alpha} \right)^{-1} \nonumber \\
&= \left( C_{\alpha} {\int_{0}^{1}  e^{\alpha \ln(s)} d \alpha} \right)^{-1}
 = \frac{1}{C_{\alpha}} \frac{\ln(s)}{s-1}
\label{eq:z}
\end{align}
This result proves that the parallel association of CPEs with distributed orders following a uniform distribution \emph {does not} yield an equivalent CPE,  as expected \cite{said2019modulating}. 
 The impedance given by Eq.\;\ref{eq:z} is a particular case of the more general form\;\cite{lorenzo2002variable}:
\begin{equation}
\tilde{z}_{\beta}(s)=  \frac{1}{C_{\alpha}} \frac{\ln(s)}{s^{\beta_2}-s^{\beta_1}}
\label{eq:z4}
\end{equation} 
The real and imaginary parts of $\tilde{z}_{\beta}(s)$ are  given respectively by:
\begin{align}
\text{Re}(\tilde{z}_{\beta}(s))
&=-\frac{\omega ^a \left[2 \cos \left(\frac{\pi  a}{2}\right) \ln (\omega )+\pi  \sin \left(\frac{\pi  a}{2}\right)\right]}{D(\omega,a,b)}\nonumber \\ 
&+\frac{\omega ^b \left[2 \cos \left(\frac{\pi  b}{2}\right) \ln (\omega )+\pi  \sin \left(\frac{\pi  b}{2}\right)\right]}{D(\omega,a,b)} \\
\text{Im}(\tilde{z}_{\beta}(s))&= \frac{\omega ^a \left[2 \sin \left(\frac{\pi  a}{2}\right) \ln(\omega )-\pi  \cos \left(\frac{\pi  a}{2}\right)\right]}{D(\omega,a,b)} \nonumber \\
& + \frac{\omega ^b \left[\pi  \cos \left(\frac{\pi  b}{2}\right)-2 \sin \left(\frac{\pi  b}{2}\right) \ln(\omega )\right]}{D(\omega,a,b)} \\
\text{where} \\ D(\omega,a,b)&= {2 \left[-2 \omega ^{a+b} \cos \left( \pi  (a-b)/2\right)+\omega ^{2 a}+\omega ^{2 b}\right]} \nonumber
\end{align}
clearly different from those of a single CPE  mentioned above:  $\text{Re}(z_c) = \cos(\alpha \pi/2)/(C_{\alpha} \omega^{\alpha})$, and   $\text{Im}(z_c) = -\sin(\alpha \pi/2)/(C_{\alpha} \omega^{\alpha})$. 

\begin{figure}[!t]
\begin{center}
\includegraphics[width=0.37\textwidth]{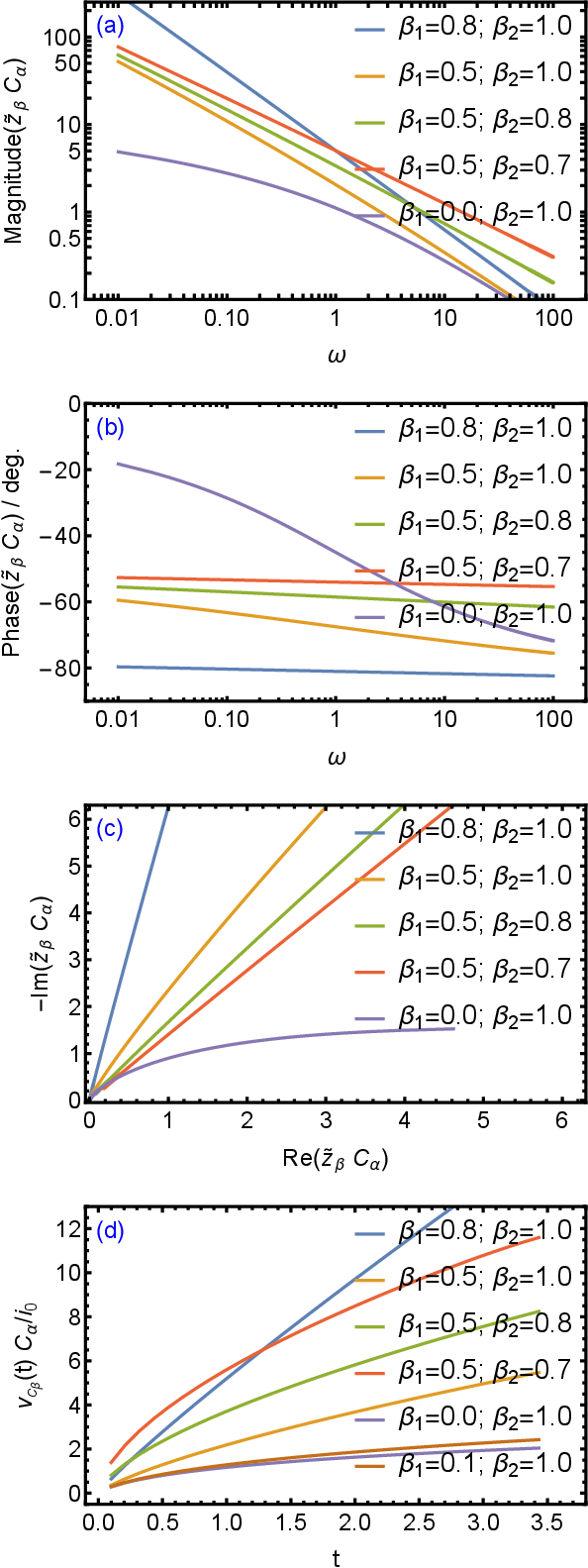}
\caption{(a)-(c):  Plots of the impedance function given by Eq.\;\ref{eq:z4}, and (d) plots  of $v_{c_{\beta}}$ (Eq.\;\ref{eq:vc4}) and its special case $v_{c_u}$ (Eq.\;\ref{eq:vct} (valid only at $\beta_1=0,\,\beta_2=1.0$))
}
\label{fig3}
\end{center}
\end{figure}

In Fig.\;\ref{fig3} we show plots of the normalized impedance function $(z_{\beta}(s) \times C_{\alpha})$   in terms of its magnitude vs. frequency (see Fig.\;\ref{fig3}(a)), its phase vs. frequency (see Fig.\;\ref{fig3}(b)) and the real vs. imaginary parts (see Fig.\;\ref{fig3}(c)) for different combination values of $\beta_1$ and $\beta_2$.  The case when $\beta_1=0$ and $\beta_2=1$, corresponding to the impedance given by Eq.\;\ref{eq:z}, is also plotted in the figures and  clearly represents the least capacitive network at low frequencies. For fixed $\beta_2=1$, the network becomes more capacitive as $\beta_1$ approaches one. We can also see from the  figures that for  the three cases where the upper limit $\beta_2$ is the same, i.e. 
($\beta_1=0.8,\,\beta_2=1.0$), 
($\beta_1=0.5,\,\beta_2=1.0$) and 
($\beta_1=0.0,\,\beta_2=1.0$), the impedance phase tends to the same limit at high frequencies dictated by this value of $\beta_2$. Whereas when the lower limit $\beta_1$ is the same, i.e. 
($\beta_1=0.5,\,\beta_2=1.0$), 
($\beta_1=0.5,\,\beta_2=0.8$) and 
($\beta_1=0.5,\,\beta_2=0.7$),  the   phase tends to the same limit at low frequencies. 
Otherwise, it is evident that with this particular superposition of CPEs ($\phi(\alpha)$ uniform distribution), the resulting system is not a CPE itself.  The closest the impedance phase  gets to a constant, and therefore to a traditional CPE, is when the limits of integration $\beta_1$ and $\beta_2$ are closer to each other as it is the case for ($\beta_1=0.8,\,\beta_2=1.0$) for example.

Now we  derive the network's voltage in response to a constant current excitation  with  $i(t) = i_0>0$ for $t >0$ (see Fig.\;\ref{fig1}), assuming that the network is initially uncharged. The developed voltage $v_{c_u}(t)$ on the network  when $\beta_1=0$ and $\beta_2=1$ is given by:
\begin{align}
v_{c_u}(t)&= \mathcal{L}^{-1}\left\{ \frac{i_0}{ C_{\alpha}  } \frac{\ln(s)}{s(s-1)} ;t\right\} \\
& = \frac{i_0}{ C_{\alpha}  } \left[ \gamma - e^t  \text{Ei}(-t) + \ln(t) \right]
\label{eq:vct}
\end{align}
where $\text{Ei}(z) = - \int_{-z}^{\infty} u^{-1} e^{-u} du$ is the exponential integral function  and $\gamma$ is Euler's constant $\approx 0.577216$. Similar results were reported by Eab and Lim  \cite{eab2011fractional} for the case of simple free fractional Langevin equation without a friction  term.

The voltage expression for the general impedance function under the condition $0<\beta_1\leqslant \alpha \leqslant \beta_2<1$  can also be derived after applying the  series expansion $(1-x)^{-1}=\sum_{k=0}^{\infty}  x^k$ for $|x|<1$ enabling us to write:
\begin{align}
v_{c_{\beta}}(t)&= \mathcal{L}^{-1}\left\{ \frac{i_0}{ C_{\alpha}  } \frac{ s^{-\beta_2-1} \ln(s)}{1-s^{\beta_1-\beta_2}} ;t\right\} \\
 & = \frac{i_0}{ C_{\alpha}} \mathcal{L}^{-1}\left\{    \ln(s)   \sum_{k=0}^{\infty}   s^{-k(\beta_2-\beta_1)-(\beta_2+1)} ;t\right\}
 \label{eq:vct2}
\end{align}
Noting that the $\ln(s)$ function can be represented in terms of the Fox $H$-function as: 
\begin{align}
\ln(s) &= \ln (1+s) - \ln(1+s^{-1}) \nonumber \\
 &=  
H^{1,2}_{2,2}\left[ s  \left|
\begin{array}{l}
(1,1), (1,1)    \\
(1,1), (0,1)     \\
\end{array}
\right.\right]-  
H^{1,2}_{2,2}\left[ s^{-1}  \left|
\begin{array}{l}
(1,1), (1,1)    \\
(1,1), (0,1)     \\
\end{array}
\right.\right] 
\end{align} 
the inverse Laplace transform can then be evaluated term by term.  
Recall that the $H$-function  of order 
$(m,n,p,q)\in \mathbb{N}^4$, ($0 \leqslant n \leqslant p$, $1 \leqslant m \leqslant q$) 
and with parameters 
  $A_j \in \mathbb{R}_+ \;(j=1,\ldots,p)$, $B_j \in \mathbb{R}_+\; (j=1,\ldots,q) $, 
$a_j \in\mathbb{C}\;(j=1,\ldots,p)$ and $ b_j \in \mathbb{C}\; (j=1,\ldots,q)$ 
is defined for $z \in \mathbb{C},\;z\neq 0$ by the contour  integral\;\cite{mathai2009h, mathai1978h}:
\begin{equation}
H^{m,n}_{p,q}\left[ z\left|
\begin{array}{c}
{(a_1,A_1),\ldots,(a_p,A_p)}\\
{(b_1,B_1),\ldots, (b_q,B_q)}
\end{array} 
\right.\right]
=\frac{1}{2\pi i} \int_L h(s) z^{-s} ds
\label{eq:H}
\end{equation}
 where the integrand $h(s)$ is given by:
 \begin{equation}
h(s) = \frac{\left\{\prod\limits_{j=1}^m \Gamma(b_j + B_j s)\right\}  \left\{\prod\limits_{j=1}^n \Gamma(1-a_j - A_j s)\right\}}
{\left\{\prod\limits_{j={m+1}}^q \Gamma(1-b_j - B_j s)\right\} \left\{\prod\limits_{j={n+1}}^p \Gamma(a_j + A_j s)\right\}}
\end{equation}
In Eq.\;\ref{eq:H}, $z^{-s}=\exp \left[ -s (\ln|z|+ i \arg z) \right] $ and $\arg z$ is not necessarily the principal value. 
 The contour of integration $L$ is a suitable contour separating the poles  
 of  $\Gamma(b_j+ B_j s)$ ($j=1,\ldots,m$) from the poles 
 of   $\Gamma (1-a_{j} - A_{j} s)$ ($j=1,\ldots,n$).
 An empty product is always interpreted as unity. 
 We also recall that the Laplace transform of an $H$-function which   is given by \cite{mathai2009h}: 
\begin{align}
\label{eq:LTH}
\mathcal{L} 
&\left[ 
t^{\rho-1} H_{p,q}^{m,n}\left[ a t^{\sigma} \left|
\begin{array}{l}
(a_p,A_p)  \\
(b_q,B_q)  \\
\end{array}
\right.\right];  u
\right] \\ \nonumber
 &= 
  u^{-\rho} 
H^{m,n+1}_{p+1,q}\left[ a  u^{-\sigma}\left|
\begin{array}{l}
(1-\rho,\sigma), (a_1,A_1), \ldots, (a_p,A_p)   \\
(b_1,B_1), \ldots, (b_q,B_q) \hfill \\
\end{array}
\right.\right]
\end{align}
 and the inverse Laplace transform by \cite{mathai2009h}:
\begin{align}
\mathcal{L}^{-1} 
&\left[ 
  u^{-\rho} 
H_{p,q}^{m,n}\left[ a  u^{\sigma}\left|
\begin{array}{l}
(a_p,A_p)  \\
(b_q,B_q)  \\
\end{array}
\right.\right]; t
\right] \label{eq:ILT} \\ \nonumber
 &= t^{\rho-1} 
H_{p+1,q}^{m,n}\left[ a t^{-\sigma}\left|
\begin{array}{l}
(a_p,A_p), \ldots, (a_1,A_1), (\rho,\sigma)  \\
(b_1,B_1),\ldots,(b_q,B_q) \hfill   
\end{array}
\right.\right]
\end{align} 
Thus we obtain the general solution for  $v_{c_{\beta}}(t)$ given by Eq.\;\ref{eq:vct2} as:
\begin{align}
v_{c_{\beta}}(t)= \frac{i_0}{ C_{\alpha}} 
\sum_{k=0}^{\infty}  t^{\rho_k-1}     
&\left\{ H^{1,2}_{3,2}\left[ t^{-1}  \left|
\begin{array}{l}
(1,1), (1,1) , ( {\rho_k},1)   \\
(1,1), (0,1)     
\end{array}
\right.\right] \right.  \nonumber \\ 
& \left.  - H^{1,2}_{3,2} \left[ t  \left| 
\begin{array}{l}
(1,1), (1,1) , ( {\rho_k},-1)   \\
(1,1), (0,1)     
\end{array}
\right.\right] \right\},
\label{eq:vc4}
\end{align} 
where $\rho_k=k(\beta_2-\beta_1)+(\beta_2+1)$. This represents the total voltage developed across the network with the only constraint being that $0<\beta_1\leqslant \alpha \leqslant \beta_2<1$.

In Fig.\;\ref{fig3}(d) we plot the normalized network voltage $(v_{c_{\beta}}(t) \times C_{\alpha}/i_0)$  as given by Eq.\;\ref{eq:vc4} for different values of $\beta_{1}$ and $\beta_{2}$. The summation was truncated to the first 40 terms. Also,  the normalized network voltage  $(v_{c_u}(t) \times C_{\alpha}/i_0)$ given by Eq.\;\ref{eq:vct} (i.e. when $\beta_1=0.0$ and $\beta_1=1.0$) is plotted in the same figure. Correspondingly, a plot is shown for the case of $\beta_1=0.1,\,\beta_2=1.0$  in order to compare the accuracy of plotting the $H$-function based expression vs.  directly plotting Eq.\;\ref{eq:vct} (valid only for $\beta_1=0.0,\,\beta_2=1.0$). It can be clearly seen  that in both these cases, the steady-state normalized voltage (i.e. at $t \rightarrow \infty$) has the lowest value  because the network is least capacitive, as expected. The higher the value of the steady-state voltage, the more capacitive the network is. 
 In particular, at a first glance when comparing the voltage profiles simulated with  
($\beta_1=0.8,\,\beta_2=1.0$), 
($\beta_1=0.5,\,\beta_2=1.0$) and 
($\beta_1=0.0,\,\beta_2=1.0$) having the same values for $\beta_2$, it appears  that the more the interval $[\beta_1,\beta_2]$ is skewed to one, the higher is the magnitude of the voltage. But for  
($\beta_1=0.5,\,\beta_2=1.0$), 
($\beta_1=0.5,\,\beta_2=0.8$) and 
($\beta_1=0.5,\,\beta_2=0.7$), which are all having the same lower limit $\beta_1$, the voltage magnitude seems to be higher as the value of $\beta_2 $ approaches that of $\beta_1$. The same remark now holds for fixed value of $\beta_2$ just discussed previously. But if we compare the two cases of ($\beta_1=0.8,\,\beta_2=1.0$) and 
($\beta_1=0.5,\,\beta_2=0.7$), wherein the size of the interval for $\alpha$ is 0.2 for both, we can clearly see a crossing point below which the latter case generate higher voltage, and vice versa, i.e. above this crossing point it is the voltage corresponding to ($\beta_1=0.8,\,\beta_2=1.0$) that overtakes the other.  This is an interesting observation that highlights the difference in response time of distributed order CPEs at small vs. larger values of time.

\subsection{Example 2: $\phi(\alpha)$ is a set of discrete values}

As a second example, we   consider the case where the weight function for the order $\alpha$ is given by:
\begin{equation}
\phi(\alpha) = \sum_i C_{\alpha_i}\, \delta(\alpha-\alpha_i)
\end{equation}
 with $\delta(x)$  the Dirac delta function. 
 

The network's equivalent impedance is straightforward to obtain in this case. For example, for $i=2$, the network contains only two CPEs with different parameters, and  its equivalent impedance is simply:
\begin{align}
\tilde{z}_2(s) &= \frac{1}{ C_{\alpha_1} s^{\alpha_1} + C_{\alpha_2} s^{\alpha_2} } \label{eq:z2} 
\end{align} 
This means that  we have a total current in the system being:
\begin{equation}
i_2(t) = 
C_{\alpha_1}\, {}_0\text{D}_t^{\alpha_1} v_c(t) + 
C_{\alpha_2}\, {}_0\text{D}_t^{\alpha_2} v_c(t)
 \end{equation}
 and therefore the order distribution function $\phi(\alpha)$  is:
\begin{equation}
\phi_2(\alpha) = C_{\alpha_1}\, \delta(\alpha-\alpha_1) + 
C_{\alpha_2}\, \delta(\alpha-\alpha_2)
\end{equation} 
While for $i=3$, i.e. the case of three CPEs in parallel the equivalent impedance  is:
\begin{align}
\tilde{z}_3(s) = \frac{1}{ C_{\alpha_1} s^{\alpha_1} + C_{\alpha_2} s^{\alpha_2} + C_{\alpha_3} s^{\alpha_3} } \label{eq:z3} 
\end{align} 
and for the general case, the total network input admittance is the sum:
\begin{align}
\tilde{y}_i(s) &={ C_{\alpha_1} s^{\alpha_1} + C_{\alpha_2} s^{\alpha_2} }+.....+C_{\alpha_i} s^{\alpha_i} 
\end{align} 
   
\begin{figure}[!th]
\begin{center}
\includegraphics[width=0.37\textwidth]{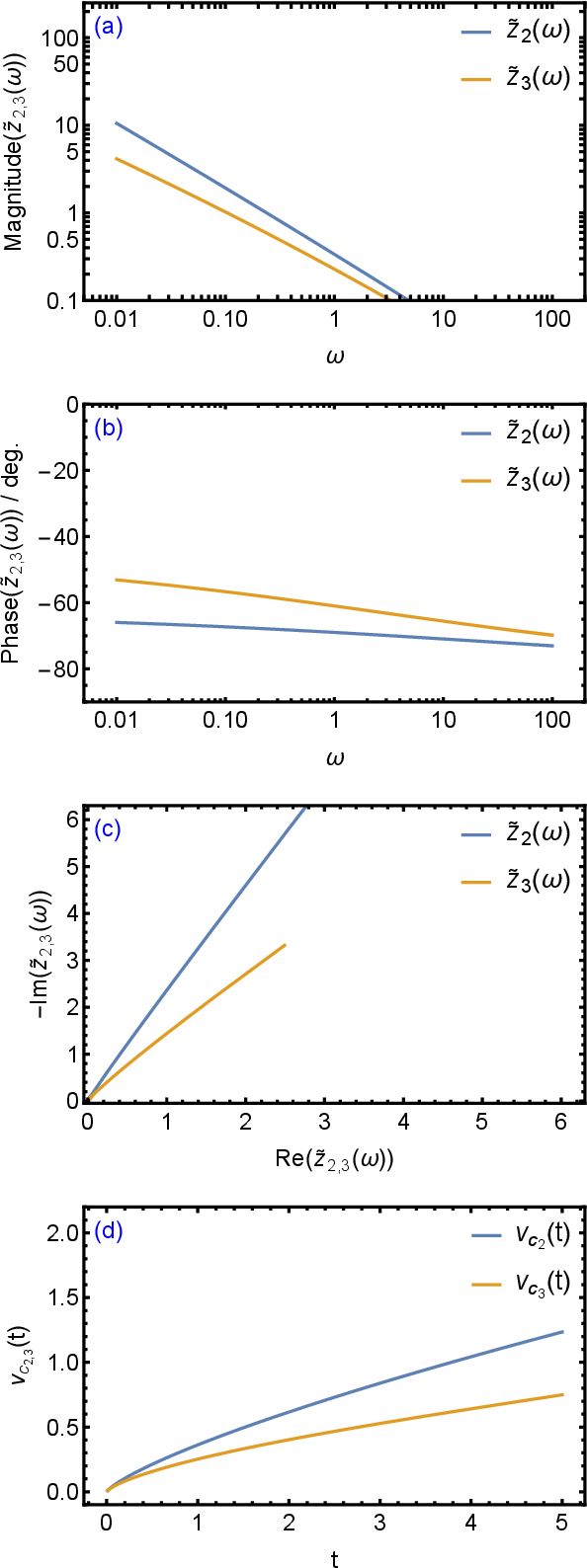}
\caption{(a)-(c):  Plots of the impedance functions $\tilde{z}_2(s)$ (Eq.\;\ref{eq:z2}) and $\tilde{z}_3(s)$ (Eq.\;\ref{eq:z3}), and (d) plots of the voltages $v_{c_2}(t)$ (Eq.\;\ref{eq:vc2}) $v_{c_3}(t)$ (Eq.\;\ref{eq:vc3}) }
\label{fig2}
\end{center}
\end{figure}

   For illustration purposes, we show in Fig.\;\ref{fig2}  plots of $\tilde{z}_2(s)$ (Eq.\;\ref{eq:z2}) and $\tilde{z}_3(s)$ (Eq.\;\ref{eq:z3})  in terms of magnitude vs. frequency (Fig.\;\ref{fig2}(a)), phase vs. frequency (Fig.\;\ref{fig2}(b)) and real vs. imaginary parts (Fig.\;\ref{fig2}(c)) with the parameters values  
 ($\alpha_1=0.9$, $C_{\alpha_1}=1.0\,\text{F\,s}^{\alpha_1-1}$), 
 ($\alpha_2=0.9$, $C_{\alpha_2}=2.0\,\text{F\,s}^{\alpha_2-1}$), 
 ($\alpha_3=0.5$, $C_{\alpha_3}=1.5\,\text{F\,s}^{\alpha_3-1}$), over the frequency range 0.01 to 100\,Hz. The effect of adding the third less-capacitive CPE (with $\alpha=0.5$) on the overall network impedance is clear from the figures.
 
Deriving in closed form the expression for the voltage response across the network is done here after recalling the Laplace transform formula \cite{prabhakar1971singular, saxena2004generalized}:
\begin{equation}
  \int \limits_0^{\infty}  t^{\beta-1}  {E}_{\alpha,\beta}^{\gamma} \left( -at^{\alpha}\right) e^{-st} dt =  \frac{s^{-\beta}}{(1+as^{-\alpha})^{\gamma}} 
  \label{e5}
\end{equation}
where 
\begin{equation}
\text{E}_{\alpha,\beta}^{\gamma} ( z ) := \sum\limits_{k=0}^{\infty} \frac{(\gamma)_k}{\Gamma(\alpha k + \beta)} \frac{z^k}{k!} \quad (\alpha,\beta, \gamma \in \mathbb{C}, \mathrm{Re}({\alpha})>0)
\label{eqML}
\end{equation}
  is the generalized Mittag-Leffler function  \cite{prabhakar1971singular} and $(\gamma)_k=\Gamma(\gamma+k)/\Gamma(\gamma)$ is the Pochhammer symbol. We can  obtain directly the time-domain voltage in response to a constant current excitation  $i(t) = i_0>0$ for $t >0$ applied on the (initially uncharged) network with two CPEs only as: 
\begin{equation}
v_{c_2}(t) = C_{\alpha_1}^{-1}{i_0}\,  t^{\alpha_1} 
\text{E}_{\alpha_1-\alpha_2,\alpha_1+1} ( - C_{\alpha_1}^{-1}  C_{\alpha_2} t^{\alpha_1-\alpha_2} )
\label{eq:vc2}
\end{equation}
Note that the Mittag-Leffler function can also be expressed as an $H$-function\;\cite{mathai2009h} in the form:
\begin{equation}
\text{E}_{\alpha,\beta}^{\gamma} ( z ) = 
\frac{1}{ \Gamma(\gamma) } 
H^{1,1}_{1,2}\left[ -z  \left|
\begin{array}{l}
(1-\gamma,1)    \\
(0,1), (1-\beta, \alpha)     \\
\end{array}
\right.\right] 
\end{equation} 
We verify that when $\alpha_1=\alpha_2$ in Eq.\;\ref{eq:vc2},  {the voltage simplifies to:} 
\begin{equation}
v_{c_2}(t) = \frac{{i_0}\,  t^{\alpha_1}}{  (C_{\alpha_1}+C_{\alpha_2}) \Gamma(1+ \alpha_1)}
\end{equation}
and if we additionally have $C_{\alpha_1}=C_{\alpha_2}$, then 
\begin{equation}
v_{c_2}(t) = \frac{{i_0}\,  t^{\alpha_1}}{   2 C_{\alpha_1}  \Gamma(1+ \alpha_1)}
\end{equation}
as expected.

Now for the case of three CPEs in the network and under the assumption that $\alpha_1>\alpha_2 >\alpha_3$, it can be shown  that the network equivalent impedance given by Eq.\;\ref{eq:z3} can be rewritten as:
\begin{align}
\tilde{z}_3(s) 
 &= \cfrac{ C_{\alpha_1}^{-1} s^{-\alpha_1}}{ 1+ C_{\alpha_1}^{-1} C_{\alpha_2} s^{\alpha_2-\alpha_1}  } 
\cfrac{1}{1+ \cfrac{ C_{\alpha_1}^{-1} C_{\alpha_3} s^{\alpha_3-\alpha_1} }{ 1+ C_{\alpha_1}^{-1} C_{\alpha_2} s^{\alpha_2-\alpha_1}   }
}
 \\
&=  C_{\alpha_1}^{-1} \sum\limits_{k=0}^{\infty} \left( - C_{\alpha_1}^{-1} C_{\alpha_3}  \right)^{k} \cfrac{ s^{-\alpha_1 + k(\alpha_3-\alpha_1)} }{ \left(1 + C_{\alpha_1}^{-1}  C_{\alpha_2} s^{\alpha_2 - \alpha_1} \right)^{k+1} }
\end{align} 
with  the use of the expansion formula   $(1+x)^{-1}=\sum_{k=0}^{\infty} (-1)^k x^k$ for $|x|<1$. 
Its inverse Laplace transform is carried out term by term using Eq.\;\ref{e5} leading to:
\begin{align}
z_3(t)=   C_{\alpha_1}^{-1} t^{\alpha_1 -1} \sum\limits_{k=0}^{\infty} & \left( - C_{\alpha_1}^{-1}  C_{\alpha_3}  \right)^{k}  
t^{{ - k(\alpha_3-\alpha_1)}} \nonumber \\
& \times \text{E}_{\alpha_1-\alpha_2,{\alpha_1 - k(\alpha_3-\alpha_1)}}^{k+1} \left( - C_{\alpha_1}^{-1}  C_{\alpha_2} t^{\alpha_1-\alpha_2}\right)
\end{align}
Under a constant current excitation, the resulting voltage for this case is found to be:
\begin{align}
v_{c_3}(t)=   i_0 C_{\alpha_1}^{-1} t^{\alpha_1} \sum\limits_{k=0}^{\infty} & \left( - C_{\alpha_1}^{-1} C_{\alpha_3}  \right)^{k} 
t^{{ - k(\alpha_3-\alpha_1)}} \nonumber \\
& \times \text{E}_{\alpha_1-\alpha_2,{\alpha_1 +1 - k(\alpha_3-\alpha_1)}}^{k+1} \left( - C_{\alpha_1}^{-1}  C_{\alpha_2} t^{\alpha_1-\alpha_2}\right)
\label{eq:vc3}
\end{align}


The solution to a similar problem  for $n$ discrete terms as depicted in Fig.\;\ref{fig1}, i.e. for a total current of the form:
\begin{equation}
i_n(t) = 
C_{\alpha_1}\, {}_0\text{D}_t^{\alpha_1} v_c(t) + 
C_{\alpha_2}\, {}_0\text{D}_t^{\alpha_2} v_c(t) + \ldots +
C_{\alpha_n}\, {}_0\text{D}_t^{\alpha_n} v_c(t)
 \end{equation}
 and thus a weight function, 
  \begin{equation}
\phi_n(\alpha) = C_{\alpha_1}\, \delta(\alpha-a_1) + 
C_{\alpha_2}\, \delta(\alpha-a_2) + \ldots + 
C_{\alpha_n}\, \delta(\alpha-a_n)
\end{equation}
 can be found in Podlubny\;\cite{podlubny1998fractional}.

 In Fig.\;\ref{fig2}(d), we plotted the time-domain voltage expressions given by Eq.\;\ref{eq:vc2} and by Eq.\;\ref{eq:vc3} (first 10 terms of the sum) using the same parameter values, and with $i_0=1$. The voltage growth follows non-exponential, power-law like profiles which is expected\;\cite{10.1149/1945-7111/ac621e, allagui2021possibility, ieeeted2}. The effect of the less capacitive CPE (with $\alpha = 0.5$) on reducing the steady-state value of the voltage is clear from the figure. Similar results can be obtained for more than three CPEs in the network.

\section{Conclusion}



In this work, we studied the behavior of a distributed order network  of CPEs modeled  by the  current-voltage relation given by Eq.\;\ref{eq:iCPE2} for two cases of the  weight function $\phi(\alpha)$ describing the variations in the CPE orders, i.e. (i) $\phi(\alpha)=1$ for $0 < \alpha < 1$ and zero otherwise, and $(ii)$ $\phi(\alpha) = \sum_i C_{\alpha_i}\, \delta(\alpha-\alpha_i)$. The weight functions considered  are solely dependent on  the order $\alpha$    without the assignment of any new variable to the problem.  While $\phi(\alpha)$ is in fact a characteristic function describing the underlying physics of the system under study, these two situations we analyzed are believed to be good enough to describe real electrode/electrolyte problems\;\cite{said2019modulating}. 
 We derived the expressions for both  the equivalent impedance function and the voltage response to a constant current excitation for these two cases of CPE networks, which yield results far different from the standard CPE response. 
    A CPE network with variable, time-dependent  CPE parameters will be the subject of a future study.

%
%



\section*{References}
 


\end{document}